\documentclass[preprint,12pt]{elsarticle}
\usepackage{amssymb}

\journal{Physics Letters B}

\begin{document}

\begin{frontmatter}

\title{The Photon Sector in the Quantum  Myers-Pospelov
Model: an improved description}

\author{C. M. Reyes, L. F. Urrutia and J. D. Vergara}

\address{Instituto de Ciencias Nucleares, Universidad Nacional Aut{\'o}noma de M{\'e}%
xico, \\
A. Postal 70-543, 04510 M{\'e}xico D.F., M{\'e}xico}

\begin{abstract}
The quantization of the electromagnetic sector of the Myers-Pospelov
model coupled to standard fermions is studied. Our main objective is
to construct an effective quantum theory that results in a genuine
perturbation of QED, such that setting zero the Lorentz invariance
violation (LIV) parameters will reproduce it.  This is achieved by
introducing an additional low  energy scale $M$, together with a
physically motivated prescription to take the QED limit. The
prescription is successfully tested in the calculation of the
electron self-energy in the one loop approximation. 
The LIV radiative corrections turn out to be
properly scaled by very small factors for any reasonable values of
the parameters, no fine-tuning problems are found at this stage  and the choice for
$M$ to be of the order of the electroweak symmetry breaking scale is
consistent with the stringent bounds for the LIV parameters, in
particular with those arising from induced dimension three
operators.
\end{abstract}


\end{frontmatter}

\section{Introduction}

Recently there has been a great deal  of interest in the study of
effective field theory models that describe violations of Lorentz
and CPT invariance \cite{KOSTELECKY0}  in order to correlate the
numerous and diverse experimental and observational test carried to
probe those symmetries \cite{KREV}. Such interest has been enhanced
from the theoretical perspective since the proposal of Ref.
\cite{GEMN} suggesting that Lorentz invariance violation (LIV) could
arise due to a foamy or granular structure of space-time.  Observation of
high energy photons arriving from astrophysical sources was also
proposed as a method to test such possibility \cite{PREVEELLIS} .
The above suggestion sparkled immediate interest in identifying
fundamental theories that could generate these effects. The most
natural choice to look for is a dynamical theory of space-time at the
quantum level, that is to say quantum gravity, where most of the
developing theories share the belief that the description of space-time
will suffer important deviations from its  standard view as a
continuum, when we are in the Planck scale regime. Preliminary
estimations of the induced corrections in particle propagation at
standard model energies appeared in  Refs.
\cite{GP,AMU1,AMU3,ELLIS1}. Nevertheless, up to now there is no
systematic derivation of any semiclassical approximation starting
from a fundamental quantum gravity theory, for example, that could
determine the exact nature of the possible corrections, if any,  arising from
such modifications of space-time. This situation has prompted the construction
and analysis of effective field theory models which capture the
basic ingredients that we expect to survive at standard model
energies. At present, all observational test of LIV lead to
negligible violation, codified in the very stringent bounds set upon
the LIV parameters. In this way, the proposed effective models have
to provide highly suppressed radiative corrections to comply with
observation \cite{ALFARO}. Radiative corrections to LIV
theories have been also considered in Refs. \cite{RADCORR}. Fine
tuning problems arising from LIV theories have been found in Refs.
\cite{COLLINS} and \cite{GV}. This last reference deals  with the
Myers-Pospelov model (MPM), to be discussed in this Letter from a
different perspective which eliminates those fine tuning problems.
The MPM  \cite{MP} incorporates particle (active) LIV
parameterized by dimension five operators together with a
non-dynamical timelike four-vector $n_\mu$ that can be interpreted
as the four velocity of a  preferred frame. It respects observer
(passive) Lorentz covariance among concordant frames.
Due to the presence of  $n^{\mu}$ the full MPM exhibits higher order
derivative (HOD) corrections in the kinetic terms of the Lagrangian
entering as dimension five operators. As it is well known such theories
have additional degrees of freedom with respect to the standard ones,
present unitarity and causality violations together with unbounded
Hamiltonians which are non-analytical in the coefficients $\xi_A $
that control the higher dimensional operators. Many different
approaches to cope with such problems in the Lorentz covariant
case have appeared in Refs. \cite{HOTD} and a final answer has
not been provided yet. Of particular interest to the case of
the full MPM considered  as a perturbation of QED is the work
in Ref. \cite{COREANOS}, where a consistent perturbation procedure
in terms of the HOD operators is developed, that allows to calculate
corrections to the physics of the original low energy degrees of
freedom ( i. e. those corresponding to standard QED in our case).
When dealing with the full MPM as a perturbation of QED, such
drawbacks must be taken carefully into account with the additional
requirement of extending the analysis to the LIV case.

In the present work we introduce some simplifications that make the
quantization of the electromagnetic sector of the MPM much simpler, without
loosing some of the general features of
the complete model. In the first place we retain only  the LIV parameter
associated to  the photon field. In
this way our starting point is the Lagrangian
\begin{eqnarray}
\mathcal{L}=-\frac{1}{4}F_{\mu \nu }F^{\mu \nu }+\frac{\xi
}{2 \bar{M}}\left( n^{\mu }F_{\mu \nu }\right) \left( n^{\alpha
}\partial _{\alpha }\right) \left( n_{\rho
}\epsilon^{\rho\nu\kappa\lambda}F_{\kappa\lambda}\right) \nonumber && \\
+\bar{\Psi}\left(i\gamma ^{\mu } (\partial _{\mu }+ieA_\mu)-m\right)
\Psi,  && \label{LAGMP}
\end{eqnarray}
where LIV is  codified by the parameter $\xi$ and $m$ is the
electron mass. The quantity  ${\bar M}$ is assumed to arise from a
fundamental theory and determines the scale where quantum gravity
effects dominate. As a second simplification, we work in the rest
frame of the preferred system, where $n^\mu=(1, {\mathbf 0})$. In
this frame the HOD term leads to a contribution quadratic in the
field velocities, thus representing a correction over the standard
photon propagation modes with no additional degrees of freedom
involved, which can be dealt with in the standard way. The theory
described by the Lagrangian in Eq.  (\ref{LAGMP}) is  an effective
one  which is valid only up to distances of the order of $1/{\bar
M}$. Guided by our goal to recover QED in the limit $\xi \rightarrow
0$ we  introduce an additional scale $M$.  Intuitively such scale
corresponds to that entering in the  regularization procedure,  via
Pauli-Villars factors for example, required in standard QED, so that
we expect  $M << {\bar M}$. In this way we introduce it  through the
same choice of regulating factors as required in the Lorentz
invariant situation. In the present case this prescription amounts
to the following modification of the photon propagator
\begin{equation}
\Delta_{\mu\nu}(k) \longrightarrow  \Delta_{\mu\nu}(k){\cal I}(k),
\quad {\cal I}(k)=\frac{M^2}{M^2-k^2}, \quad
M >> m. \label{PV}
\end{equation}
Notice that the scale $M$ has been introduced in a fully Lorentz
covariant way such  that all LIV is still codified by the parameter
$\xi$.
Consistency with the choice of $M$ provides the prescription to recover QED from
the quantum modified model: (i) first take $ \xi
\rightarrow 0$,  for fixed $M$ and  (ii) subsequently take ${ M
\rightarrow \infty}$.  In this work we focus on the
calculation of the electron self-energy, with special emphasis upon
all LIV terms that  are good candidates to induce
fine-tuning problems by generating lower dimensional operators with
unsuppressed corrections.
\section{The quantization of  the photon sector}
To be on the safe side and motivated by the difficulties inherent to
HOD theories, some of which are still present in spite of our simplifications,
we proceed to quantize the system using a standard canonical approach,
which allows a good control over the conflicting issues.  Moreover,
taking advantage of the selected reference frame we choose to incorporate
the corrections as part of an exact free propagator which induces  modified
dispersion relations.  The canonical  approach applied to the $3+1$
description of the Lagrangian in Eq.
(\ref{LAGMP}) is analogous to the standard case in the Coulomb gauge,
except for modifications of order $g$  in the momenta canonically
conjugated to the fields $A^i$
\begin{equation}
\Pi _{i}=\frac{\partial \mathcal{L}_{\gamma }}{\partial \dot{A}^{i}}%
=\dot{A}^{i}+\partial _{i}A^{0}+2g\epsilon ^{ijk}\partial
_{j}\dot{A}^{k}, \qquad g=\frac{\xi}{\bar M}. \label{MOMENTA}
\end{equation}%
The elimination of the velocities in terms of the momenta can in fact be
performed starting from Eq. (\ref{MOMENTA}), but requires the
introduction of the non-local inverse of the operator $\left( \delta
^{ik}+2g\epsilon ^{ijk}\partial _{j}\right)$, which can be exactly
calculated.
The  canonical transformation $
\mathbf{A}_{T}\rightarrow \mathbf{\bar{A}}_{T},\;\mathbf{\Pi }%
^{T}\rightarrow \mathbf{\bar{\Pi}}^{T}$
\begin{eqnarray}
A_{T}^{i}&=&\frac{\sqrt{1+W}}{\sqrt{2}%
W}\left[ \delta ^{iq}-\frac{2g}{\left( 1+%
W\right) }\epsilon ^{imq}\partial _{m}%
\right] \bar{A}_{T}^{q}, \nonumber \\
\Pi _{r}^{T}&=&\frac{\sqrt{1+W}}{\sqrt{2}}%
\left[ \delta^{rq}+\frac{2g}{\left( 1+W%
\right) }\epsilon ^{rmq}\partial _{m}\right] \bar{\Pi}_{q}^{T},
\label{PCPI}
\end{eqnarray}%
with the notation $ W=\sqrt{ 1+4g^2\nabla^2}$ leads to the Hamiltonian
\begin{eqnarray}
H=\int d^{3}x\; \Big( \frac{1}{2}\bar{\Pi}_{p}^{T}\bar{\Pi}_{p}^{T}-%
\frac{1}{2} \bar{A}_{T}^{r} \left( \frac{\nabla ^{2}}{W^2}\right)
\left[ \delta ^{rp}-2g\epsilon
^{rnp}\partial _{n}\right] \bar{A}_{T}^{p}  &&\nonumber\\
+\frac{1}{2}J^{0}\left( -\frac{1}{%
\nabla ^{2}}\right) J^{0}-J^{i}A_{T}^{i} ({\bar A}_{T}^{k}) \Big) .&& \label{HCANFIN4}
\end{eqnarray}
exhibiting the proper normalization of the ${\bar \Pi}^2$  term.
Next we determine the corresponding normal modes of
the free ($J^\mu=0$) Hamiltonian in Eq. (\ref{HCANFIN4}) starting from the
expansion
\begin{eqnarray}
\bar{A}_{T}^{i}(x)=\int \frac{d^{3}\mathbf{k}}{\sqrt{(2\pi )^{3}}}%
\,\sum_{\lambda =\pm 1 }\sqrt{\frac{1}{2\omega _{\lambda
}(\mathbf{k)}}}\Big[
a_{\lambda }(\mathbf{k})\,\varepsilon ^{i}(\lambda ,\mathbf{k}%
)e^{-ik(\lambda )\cdot x} &&\nonumber \\
+a_{\lambda }^{\dag
}(\mathbf{k})\,\varepsilon ^{i\ast }(\lambda
,\mathbf{k})e^{+ik(\lambda )\cdot x}\Big] ,&& \label{AMUBAREXP}
\end{eqnarray}%
in terms of creation-annihilation operators $a_{\lambda }^{\dag }(\mathbf{k}%
),\; a_{\lambda }(\mathbf{k})$, respectively. The notation is $
\left[ k(\lambda )\right] _{\mu }=(\omega _{\lambda }(\mathbf{k)},-\mathbf{k}%
)$, together with $k(\lambda )\cdot x=\omega _{\lambda }(\mathbf{k)}x^{0}-\mathbf{%
k\cdot x}$, where the modified normal frequencies will be
consistently determined and
the polarization vectors $\varepsilon^{i}(\lambda ,%
\mathbf{k})$, are chosen in  the helicity
basis.  Assuming  the standard creation-annihilation commutation
rules $
[ a_{\lambda }(\mathbf{k}),\;a_{\lambda ^{\prime }}^{\dag }(\mathbf{k}%
^{\prime })] =\delta _{\lambda \lambda ^{\prime }}\delta ^{3}(\mathbf{%
k-k}^{\prime }) $ and starting from Eq. (\ref{AMUBAREXP}) we recover the
basic field commutator corresponding to the quantum mechanical
extension of the standard transverse Dirac brackets. The modified dispersion
relations are
\begin{equation}
\omega _{\lambda }^{2}\left( \mathbf{k}\right) =\frac{|\mathbf{k|}^{2}}{%
\left[ 1+2\lambda g|\mathbf{k|}\right] },  \label{RELDISFIN}
\end{equation}%
which is exact in $g$. With no loss of generality we assume from now
on that $g>0$. One can further verify that the resulting free Hamiltonian is
in fact positive definite and that has the expected expression in
terms of the previously introduced creation-annihilation operators
and the  frequencies given in Eq. (\ref{RELDISFIN}).  Also,  the
Hamiltonian is Hermitian as far as the frequencies remain real,
which is the case in the region $|{\mathbf k}|< 1/(2g)$.

Let us notice  that in Eq. (\ref{AMUBAREXP}) the four-vector
$[k(\lambda=+1)]_\mu$ is space-like, while $[k(\lambda=-1)]_\mu$ is
timelike. At this stage we are confronted with two problems that
usually arise in LIV theories: (i) on one hand, the frequency
$\omega_-({\mathbf k})$ will become imaginary  when $|\mathbf{k}|>
1/(2g)$ and diverges when $|\mathbf{k}|=|\mathbf{k}|_{max}= 1/(2g)$.
>From an intuitive point of view we consider  $1/(2g)$ as the
analogous of the value $|\mathbf{k}|_{\max }=\infty$  in the
standard case and we will cut all momentum  integrals at this value.
(ii) on the other hand, since $[k(\lambda=+1)]_\mu$ is space-like,
we can always perform an observer Lorentz transformation such that
$\omega_+({\mathbf k})$ becomes negative thus introducing stability
problems in the model. For a given momentum $\mathbf{k}$ this occurs
when  $1/\sqrt{1+2g|\mathbf{k}|}<|\mathbf{v}| <1$.  Then, the maximum allowed
momentum $|\mathbf{k}|=1/(2g)$ leads to the requirement that the allowed
concordant frames in which the quantization will remain consistent
are such that $\beta < 1/\sqrt{2}$, with respect to the rest
frame.

Next we calculate  the modified free photon propagator, given by the
standard expression $i\bar{\Delta}_{ij}(x-y) = \left\langle 0\right| T\left( {\bar{A}%
_{i}^{T}\left( x\right) \bar{A}_{j}^{T}\left( y\right) }\right)
\left| 0\right\rangle$. Nevertheless some care is required in
implementing a perturbation theory based on the Hamiltonian
(\ref{HCANFIN4}) because the interaction is described by $A_T^i$,
propagating with $\Delta_{ij}$, instead of ${\bar A}_T^i$. The
propagator $\Delta_{ij}$ is directly obtained from ${\bar
\Delta}_{ij}$ via the  canonical transformation in Eq.(\ref{PCPI}). The
further inclusion of the instantaneous Coulomb term appearing in
Eq. (\ref{HCANFIN4}), following the steps of Ref. \cite{Weinbergbook}, leads to the
four dimensional propagator
\begin{eqnarray}
\Delta _{\mu \nu }(k)=\frac{1}{((k^{2})^{2}-4g^{2}\left| \mathbf{k}%
\right|^{2}k_{0}^{4})}\Big[ -k^{2}\eta _{\mu \nu
}+2igk_{0}^{2}\epsilon
^{lmr}k_{m}\eta _{l\mu }\eta _{r\nu }  \nonumber &&\\
-\frac{4g^{2}k_{0}^{4}}{k^{2}}%
k_{l}k_{r}\delta _{\mu }^{l}\delta _{\nu
}^{r} +\frac{4g^{2}k_{0}^{4}{\left| \mathbf{k}\right|
}^{2}}{k^{2}}\eta _{0\mu }\eta _{0\nu }\Big] . && \label{PROPFIN}
\end{eqnarray}
We notice  that Eq. (\ref{PROPFIN}) is just the
propagator in the Lorentz gauge obtained from the equations of
motion arising from the Lagrangian in Eq. (\ref{LAGMP}). As our exact calculation shows,
there is no high-momentum pole arising from the denominator in the above equation.
This justifies {\it a posteriori} an expansion in powers of $g$ which should amount
to treat the LIV corrections as insertions in the original QED action.
\section{The electron self-energy}
As a first step in testing the proposed construction we consider the
electron self-energy with the dynamical
modifications introduced only via the LIV photon propagator. The
starting point is
\begin{equation}
\Sigma ^{g}(p)=-ie^{2}\int \frac{d^{4}k}{(2\pi )^{4}}\gamma ^{\mu
}\left[ \frac{(\gamma \left( p-k\right)
+m)}{((p-k)^{2}-m^{2}+i\epsilon )}\right]
\gamma ^{\nu }{\Delta}_{\mu \nu }(k)\,\, {\cal I}(k)\,\,\theta \left(%
\frac{1}{2g}-|{\mathbf k}|\right).  \label{SELFENERGY}
\end{equation}%
Next we  expand the self energy in powers of the external momentum
obtaining, up to first order,
\begin{equation}
\Sigma^{g}(p)= A I+\tilde{A}\;\gamma ^{0}\gamma ^{5}+\left( B-C\right) p^{0}\gamma _{0}
+C p^{\mu
}\gamma _{\mu } +ip^{i}\tilde{C}\; \gamma ^{j}\gamma ^{k} \epsilon
^{ijk} +O(p^{2}). \label{SIGMAEXP0}
\end{equation}
The general strategy to evaluate the required
integrals is the following. Within the region of integration
($|\mathbf{k|}<1/\left( 2g\right) $), the poles in the complex
$k_{0}$ plane  of the denominators in (\ref{SELFENERGY}) have the form $
k_{01}=\mathcal{E}({|\mathbf{k|}})-i\epsilon ,\;\; k_{02}=-\mathcal{E}({|%
\mathbf{k|}})+i\epsilon, $ with $\mathcal{E}({|\mathbf{k|}})>0$.
Here $\mathcal{E}({|\mathbf{k}|})$ stands for any of
the involved energies $\omega _{\pm }\left( \mathbf{k}%
\right)$ and $E\left( \mathbf{k}\right)=\sqrt{\mathbf{k}^2+m^2} $.
In\ this way,  it is always
possible to perform a Wick rotation to the Euclidean signature such that $%
k_{0}=ik_{4}$. Due to the remaining rotational symmetry, together
with the symmetrical integration over $\mathbf{k}$, one is finally
left with only two integration variables which are $k_{4}$ and
$|\mathbf{k}|$ that  can be conveniently rewritten in polar form
$k_{4}=r\cos \alpha, \,\, {|\mathbf{k}|}=r\sin \alpha$. The details of the exact
calculation
are given in Refs. \cite{MUV0,MUV}. The potentially dangerous contributions
arise from the following terms
\begin{eqnarray}
B-C = \frac{e^{2}}{\pi ^{2}} (gM)^{2}\Big( -0.070+0.010\ln (gM)\Big)
+ \dots,\,\, &&\label{BMC}\\
\tilde{A}=\frac{e^{2}}{6\pi ^{2}}g  M^{2}\Big( 0.018+0.063\ln
(gM)\Big) + \dots, \,\, && \label{TA} \\
\tilde{C}=\frac{e^{2}}{48\pi ^{2}}\left( gm\right)\, \ln \left(
\frac{m}{M}\right) + \dots, \,\, && \label{TC} \\
A=\frac{e^{2}m}{\pi ^{2}}\Bigg( \frac{M^{2}}{2(m^{2}-M^{2})}\ln \left(
\frac{M}{m}\right) +(gM)^{2}\Big( 0.75+0.047\ln (gM)\Big)
 \Bigg) + \dots,  \,\,&&\label{A}
\end{eqnarray}
where we have written only the dominant parts.
\section{Final comments}
Contrary to the case of the second Ref. \cite{MP} we admit here the
appearance of induced lower dimensional operators. In order to make
some numerical estimations we take $\bar{M}=M_{P}=10^{19}$ GeV.  Let
us begin with the discussion of the term ${\tilde{A}}$, that will
provide an improved interpretation of  $M$ as a low energy scale,
which we take to be  the electroweak symmetry breaking  scale, i. e.
$M\simeq 250$ GeV. This point was not fully
addressed in Ref. \cite{MUV}, that was mainly concerned with the $%
\xi \rightarrow 0$ limit of the MPM. In order to take properly into
account the radiative  correction  induced by $\tilde{A}$ we have to
carry on part of the renormalization process related to the bare
coupling $\xi$. First we rewrite the photon modification term in
Eq.(\ref{LAGMP}) in terms of the renormalized coupling $\xi _{R}$ by
introducing $\xi =\xi _{R}+\xi _{R}\left( \xi /\xi _{R}-1\right) $
and subsequently we treat the second contribution arising from the
previous splitting  as a counterterm. In this way we have to change
$\xi \rightarrow \xi _{R}$  in all results of the previous section.
We take the upper bound $\xi_R =10^{-10}$, given in Ref.
\cite{LIBERATI}. Also we denote the tuning coefficient $\left( \xi
/\xi _{R}-1\right) =\mu $.  From Eq. (\ref{SIGMAEXP0}) we realize
that the radiative correction proportional to $\tilde{A}$ gives rise
to the dimension three
operator  $\left( \Delta {\cal L}\right) _{RC}=b_{0}(\xi _{R})\left[ {\bar{%
\psi}}\gamma ^{0}\gamma ^{5}\psi \right] $, dominated by  $b_{0}(\xi
_{R})=0.063\,(e^{2}/6\pi ^{2})\,\left( \xi _{R}M^{2}/\bar{M}\right) \,\ln
\left( \xi _{R}M/\bar{M}\right) $. This means that we had better started
with the corresponding bare term in the original Lagrangian (\ref{LAGMP}), which we
write in the analogous form $\left( \Delta {\cal L}\right) _{BARE}=-b_{0}(\xi )%
\left[ {\bar{\psi}}\gamma ^{0}\gamma ^{5}\psi \right]$. In this way we
obtain  $\left( b_{0}\right) _{EXP}=b_{0}(\xi _{R})-b_{0}(\xi )$ for the observable
prediction of such coefficient. Under the approximation $(1+\mu )\ln
(1+\mu )\simeq \mu $, we have
$|b_{0}|_{EXP}=\mu \times 0.063\,(e^{2}/6\pi ^{2})\,\left( \xi _{R}{M^{2}}/{%
\bar{M}}\right) \left| \,\left[ \,\,1+\,ln\xi _{R}+ln\left({M}/{\bar{M}}%
\right) \right] \right|$. The bound $|b_{0}|_{EXP}<10^{-29}$ GeV
\cite{TABLEKOSTELECKY}  leads to $\mu < 3.4\times 10^{-2}$,  which
we consider acceptable. The absence of the additional scale $M$
would lead to a tuning coefficient regulated only by $\bar M$, with
a value of $\mu\approx 10^{-34}$. Regarding the $(B-C)$
contribution, we observe that the use of any covariant regulator
$F(k^{2}/M^{2})$ to introduce the scale $M$ leads to a zero value
for the finite $g^{2}$ independent piece, thus eliminating the large
unsuppressed corrections reported in Ref.\cite{GV}. We verify that
the remaining contribution is consistent with recent observations.
In our specific case, this LIV contribution produces an additional
dimension four
term in the Lagrangian, given by $(\Delta {\cal L})_{2}=({e^{2}}/{\pi ^{2}}%
)\,\delta \,\bar{\Psi}\gamma ^{0}i\partial _{0}\Psi $,  where our
calculation leads to a prediction dominated by $|\delta |\sim
10^{-2}\times (\xi_R M/{\bar M})^2|\ln (\xi_R M/{\bar M})|=3.8\times
10^{-54}$, which falls comfortably within the observational range
$|\delta |<10^{-21}$ established in  Ref.\cite{GV}. The term
proportional to $\tilde{C}$  provides a contribution $%
(\Delta {\cal L})_{3}\sim 2\,\tilde{C}\,\bar{\psi}\gamma ^{0}\gamma ^{5}(\gamma
^{k}i\partial _{k})\psi =2\tilde{C}m\;\bar{\psi}\gamma ^{0}\gamma ^{5}\psi ,$
where, for the sake of an estimation, we have used the zeroth-order equation
of motion for $\psi $, together with dropping a remaining total time
derivative term. Again, this corresponds to a dimension three operator with $%
|b_{0}^{\prime \prime }|=2\tilde{C}m= 10^{-39}$ GeV. Finally, the
term $A$ is unsuppressed but induces a contribution to the Lorentz
covariant fermion mass term, which should be dealt with via the
fermion mass renormalization procedure. The remaining LIV
contributions to the electron self-energy given in
Eq.(\ref{SIGMAEXP0}), including corrections up to second order in
the external momentum, are calculated in analogous way and
produce highly suppressed corrections of similar type, as shown in Ref.\cite%
{MUV}. These results, with the exception of $A$, have precisely the
expected property that reduce to zero when we turn off the LIV
correction parameterized by $\xi_R $, keeping $M$ fixed. In this
letter we have presented the construction of a sector of the quantum
MP effective model emphasizing the recovering of the correct QED
limit in relation with the absence of fine-tuning problems. A low
energy scale $M \approx 250$ GeV  has been introduced which is
consistent with an ultraviolet cutoff of the order of the Planck
mass, together with the very stringent bounds upon the LIV
parameters, in particular with those associated to dimension three
operators. It is very remarkable that our procedure allows to relate
such very different UV and IR scales in a  way consistent with
observations,  including  the absence of fine tuning. C. M. R.
acknowledges support from DGAPA-UNAM through a postdoctoral
fellowship. L.F.U is partially supported by projects CONACYT \#
55310 and DGAPA-UNAM-IN109107. J. D. V acknowledges support from the
projects CONACYT \# 47211-F and DGAPA-UNAM-IN109107.

\end{document}